# Influence of cap weight on the motion of a Janus particle very near a wall

*Aidin Rashidi[1], Sepideh Razavi[2], and Christopher L. Wirth[1]*

[1]Chemical and Biomedical Engineering Department, Washkewicz College of Engineering, Cleveland State University, 2121 Euclid Ave., Cleveland, OH 44115

[2]Chemical, Biological, and Materials Engineering, University of Oklahoma, Norman, OK

**ABSTRACT**

The dynamics of anisotropic nano- to microscale 'colloidal' particles in confined environments, either near neighboring particles or boundaries, is relevant to a wide range of applications. We utilized Brownian dynamics simulations to predict the translational and rotational fluctuations of a Janus sphere with a cap of non-matching density. The presence of the cap significantly impacted the rotational dynamics of the particle as a consequence of gravitational torque at experimentally relevant conditions. Gravitational torque dominated stochastic torque for a particle > 1 μm in diameter and with a 20 nm thick gold cap. Janus particles at these conditions sampled mostly cap-down or 'quenched' orientations. Although the results summarized herein showed that particles of smaller diameter (< 1 μm) with a thin gold coating (< 5 nm) behave similar to an isotropic particle, small increases in either particle diameter or coating thickness drastically quenched the polar rotation of the particle. Histogram landscapes of the separation distance from the boundary and orientation observations of particles with larger diameters or thicker gold coatings were mostly populated with quenched configurations. Finally, the histogram landscapes were inverted to obtain the potential energy landscapes, providing a path for experimental data to be interpreted.

**I. INTRODUCTION.** Colloidal particles dispersed in a liquid interact via surface forces that play a critical role in dictating the properties and performance of complex fluid. Over the past decade, the dynamics and interactions of anisotropic colloidal particles have gained attention [1] because of potential applications in various fields such as optical displays [2], magnetorheological system [3], controlling interfacial microstructure [4], self-assembly [5,6], microfluidic devices [7], tuning interparticle interactions [8,9], and biomaterials or drug delivery [10]. Supporting these efforts have been a variety of new techniques for the synthesis of anisotropic colloidal particles [5,10–21]. Newly developed fabrication techniques provided the ability to tune the shape and the surface properties of these materials. Janus particles are one class of anisotropic colloid, typically with some property difference in the hemispherical domain. Each hemispherical domain of a Janus particle may have its own surface chemistry, shape, or other properties [22].

Predicting the dynamics of anisotropic colloids is important for applications in real systems, for example during processing when complex fluids are often not at equilibrium [23,24]. Various parameters influence the dynamics of anisotropic colloids [25–28]. Particle confinement will strongly impact the hydrodynamic interactions between the colloid and boundary, thereby strongly influencing particle mobility. Brownian motion and conservative (*i.e. path independent*) forces, such as electrostatic double layer repulsion and gravity, will also impact the dynamics of a confined spherical Janus particle. Although not dependent on orientation for an isotropic particle, each of these phenomena can be affected by the orientation of an anisotropic particle. For example, a Janus particle with anisotropy in zeta potential will experience an electrostatic interaction that depends upon orientation with respect to the boundary [29]. Rotation of the Janus particle at a constant separation distance from a boundary induces an effective change in an interaction, which then alters the probability density at that particle's position.

    Janus particles are fabricated by coating one hemisphere of a spherical colloid particle with another material, usually a metal such as gold [30,31]. The cap typically has some nominal thickness from a few to tens of nanometers, but direct measurement of the coating thickness has shown the cap to be non-uniform across the contour of the particle [32]. Tracking translational and rotational displacement of Janus particles at the various boundary, physiochemical, and rheological conditions could help with predicting the dynamics of these particles [33–36]. Various studies have focused on the rotation of isotropic [37,38] and anisotropic colloids [39,40]. There has been some work on the translational and rotational dynamics of Janus particles near a boundary [41,42] and also on the effect of mass-anisotropic coating on the dynamics of active particles away from a boundary [43,44]. Experimental techniques such as confocal microscopy [45,46], evanescent wave scattering [47], video-microscopy [48] and holographic microscopy [49–51] were used to measure the rotational diffusion coefficient. Surface roughness [52], particle shape [53], external fields [54], and the presence of motility (i.e. for active particles) [55–60] were found to influence the rotation of Janus particles. Despite the significant recent work in this area, there has not yet been a quantitative analysis of the dynamics of a Janus particle with a cap of non-matching density near a boundary.

    Herein, we conducted Brownian dynamics simulations to predict the behavior of a Janus particle with a cap of density not matching that of the native particle. This technique has been previously used to study the dynamic behavior of other colloidal systems [61–64]. Our results show the importance of the particle's surface properties, in particular, the weight of the cap, on the dynamics of a particle close to a wall. We systematically altered a variety of parameters, including

coating thickness and particle size, to test their impact on rotational and translational dynamics, probability distribution, and potential energy landscape. Our findings illustrate the importance of non-negligible torque due to either gravity or asymmetric surface forces with the substrate on the rotational and translational trajectories of a Janus particle. Notably, the aforementioned torques influence the behavior of Janus particles at conditions relevant to experimental studies.

## II. THEORY

**A.** *Near-boundary forces and torques on a Janus particle.* A colloidal particle dispersed in a Newtonian quiescent fluid near a boundary experiences conservative, dissipative, and stochastic forces. Conservative forces experienced by a non-density matched particle in low concentration electrolyte with bound surface charges are primarily due to electrostatic double layer repulsion and gravity. Strong electrostatic forces help to keep the particle levitated close to the substrate. Van der Waals attraction may also play a role at sufficiently small separation distances (h < 100 nm), but is neglected herein. A charged spherical colloidal particle with radius *a* and separation distance *h* (see **Fig. 1(a)**) will experience an electrostatic force that depends upon the size of the particle, the solution Debye length $\kappa^{-1}$, and Stern potential of the particle and the substrate. The conservative force $F_c$ is calculated by:

$$F_c = -\frac{d\phi_c(h)}{dh} = \underbrace{\kappa B \exp(-\kappa h)}_{\text{Electrostatic}} - \underbrace{\frac{4}{3}\pi a^3 (\rho_p - \rho_f) g}_{\text{Gravitational}} \quad (1)$$

$$B = 64\pi \varepsilon_0 \varepsilon_f a \left(\frac{kT}{e}\right)^2 \tanh\left(\frac{e\zeta_s}{4kT}\right) \tanh\left(\frac{e\zeta_p}{4kT}\right) \quad (2)$$

$$\kappa = \sqrt{\frac{2e^2 C_\infty}{\varepsilon_0 \varepsilon_f kT}} \quad (3)$$

where *B* is the electrostatic charge parameter, $\rho_p$ and $\rho_f$ are density of the particle and fluid respectively, *g* is gravitational acceleration, $\varepsilon_0$ is the electric permittivity of vacuum, $\varepsilon_f$ is the relative permittivity of water, *e* is the charge of an electron, $\zeta_s$ and $\zeta_p$ are the zeta potentials of the surface and particle respectively (equated with the Stern potential), *k* is the Boltzmann's constant, *T* is temperature, and $C_\infty$ is electrolyte concentration in the bulk. This expression is applicable to an isotropic particle with uniform surface chemistry. A meshing method was previously developed to account for these forces on a chemically anisotropic particle with a non-uniform zeta potential [29].

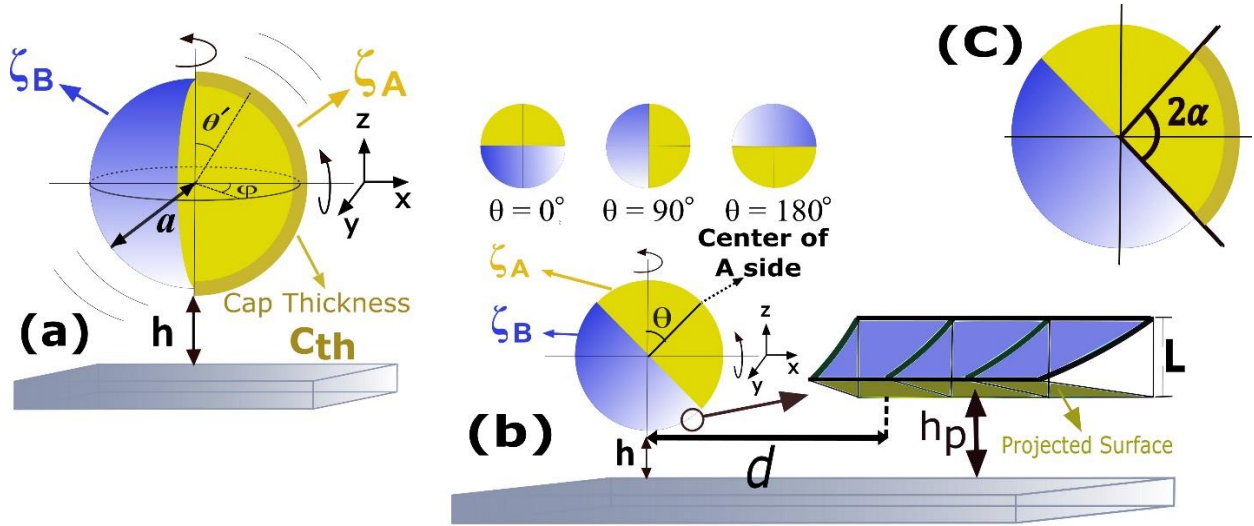

FIG. 1. (a) Schematic of a Janus particle with one hemisphere coated by 2.5 nm titanium and gold. (b) Schematic representation of θ orientation. (c) Alpha ($\alpha$) definition for the center of mass calculation.

A Janus particle will experience stochastic torque, just as an isotropic particle, but will additionally experience deterministic torque due to the asymmetry in both the gravitational and electrostatic interaction. For instance, a mismatch in the electrostatic charge on the surface of a particle induces an electrostatic torque that is important when the Janus particle's boundary is rotated downward towards the wall. Similar to accounting for variations in surface chemistry in calculating translational displacements as mentioned above, a meshing method was used to calculate the electrostatic torque on a Janus particle. For each mesh point, the torque ($T_{dL}(i)$) was equated to the product of the electrostatic force ($F_{dL}(i)$) and the distance of the projected mesh point from the center of the particle (*d*) (Fig. 1b):

$$P_{dL}(i) = 64 C_\infty kT \tanh\left(\frac{e\zeta_s}{4kT}\right) \tanh\left(\frac{e\zeta_{mp}(i)}{4kT}\right) \exp\left(-\kappa h_{dL}(i)\right) \tag{4}$$

$$T_{dL}(i) = P_{dL}(i) A(i) \times d \tag{5}$$

where $P_{dL}$ is the electrostatic double layer repulsion pressure of the substrate and projection of curved mesh area ($A(i)$), $\zeta_{mp}$ is the zeta potential of the flat projected surface of a mesh point, $h_{dL}$ is equal to substrate distance from the mid-point of the curved meshed surface ($h_{dL} = h_p + L/2$). The total torque was calculated by summing contributions from each projected flat surface area. The sum of clockwise and counter-clockwise torques at each time step provided the electrostatic torque on the Janus particle:

$$T_{dL\_total} = \sum T_{dL}(i) \tag{6}$$

where $T_{dL\_total}$ is electrostatic torque between the Janus particle and the substrate.

The gravitational torque on a Janus particle originates from the density distribution mismatch between the cap and native particle. Although gravitational deterministic torque affects the rotation of the particle about an axis parallel to the substrate, there is no impact on the rotation of the particle about the z-axis (see Fig. 1) because of the particle's axisymmetric geometry. Note that for the work summarized herein, we assumed the coating thickness distribution to be uniform over the contour of the particle. The weight of right and left hemispheres of the Janus particle was calculated at each time step with respect to a dynamic spherical coordinate system and a vertical plate that passes through the particle center. The gravitational torque was calculated:

$$T_{G\_total} = (\text{weight\_Cap}_{right} - \text{weight\_Cap}_{left}) \times CM \tag{7}$$

$$CM = \frac{2a}{\pi \times \sin(\alpha)} \times \left(\frac{\alpha}{2} + \frac{\sin(2\alpha)}{4}\right) \tag{8}$$

where $CM$ is the center of mass for (Fig. 1b) and $\alpha$ half of the angle swept out by the unmatched portion of the cap (Figs. 1b, 1c, & 8). The unmatched portion of the cap was only considered herein because the torque due to both the native particle and the matched portions of the cap cancel. Additional details regarding $CM$ calculations for the cap are found in the appendix at the end of this article.

**B.**     *Near-boundary diffusion coefficients.* Although hydrodynamic hindrance from the nearby boundary will occur in all translational (x, y, z) and orientational (θ, ϕ) directions (see Fig. 1), we are primarily concerned with hindrance in the polar rotational (θ) and translational direction normal to the substrate (z). A solution of the Stokes equation is necessary to account for the bounding effect of the wall on hydrodynamic interactions. Goldman et al. [65] provided an infinite-series solution for this equation. The normal translational diffusion coefficient of a spherical particle can be computed by an approximation of the Goldman infinite-series solution by a regression [66]:

$$D_z = \frac{kT}{f_\infty} q(h) = \frac{kT}{6\pi\eta a} q(h) \tag{8}$$

$$q(h) = \frac{6h^2 + 2ha}{6h^2 + 9ha + 2a^2} \tag{9}$$

where $T$ is temperature, $k$ is Boltzmann constant, $\eta$ is the fluid viscosity, $f_\infty$ is the friction coefficient ($f_\infty = 6\pi\eta a$), and $q(h)$ is the wall correction factor.

Dean and O'Neill [67] considered the polar rotation of a bounded sphere about an axis parallel to the planar surface. Goldman [65,68,69] updated the Dean and O'Neill solution and numerically computed the solution in form of dimensionless force and torque on a rotating sphere. We previously fit several gap ranges to implement the corrected Dean and O'Neill expression based on the numerical fit from Goldman. [29] The comprehensive polar rotational diffusion coefficients are as follows:

$$D_{r,\theta} = \frac{kT}{f_{r,\infty}}/q^\theta(h) = \frac{kT}{8\pi\eta a^3}/q^\theta(h) \tag{10}$$

$$q^\theta(h) = 0.9641\left(\frac{h}{a}\right)^{-0.1815} \qquad \frac{h}{a} \leq 0.6 \tag{11}$$

$$q^\theta(h) = 1.056\left(\frac{h}{a}\right)^{-0.07286} \qquad 0.6 \leq \frac{h}{a} \leq 2 \tag{12}$$

$$q^\theta(h) = 1 \qquad \frac{h}{a} > 2 \tag{13}$$

**C.** *Brownian dynamic simulation (BDS).* We carried out Brownian dynamics simulations to track rotational and z-axis translational motion of a Janus particle. The Langevin equation with a thermal fluctuating force, conservative forces, hydrodynamic forces, as well as torques, was used to formulate a stepping algorithm for the Janus particle. A well-known inertia-less numerical solution was obtained by Ermak and McCammon [70] to solve the Langevin equation at small Reynolds numbers. This numerical solution was used to predict the dynamic behavior of a single particle and consequently track the position of the particle at consecutive time steps. The z-axis translational and polar rotational trajectories of a single particle were predicted via Ermak and MaCammon stepping algorithm:

$$h(t + \Delta t) = h(t) + \frac{dD_z}{dh}\Delta t + \frac{D_z}{kT}F\Delta t + H(\Delta t) \tag{14}$$

$$\theta(t + \Delta t) = \theta(t) + \frac{D_{r,\theta}T_{det}\Delta t}{kT} + G(\Delta t) \tag{15}$$

These step algorithms are valid when the time step ($\Delta t$) is longer than the momentum relaxation time of the particle and is short enough such that the system properties are constant. In the z-axis translational algorithm, $F$ represents conservative forces on the particle and $H$, which has $\langle H^2 \rangle = 2D_z\Delta t$ variance, is the Gaussian random fluctuation due to Brownian motion. The total force $F$ was the sum of the double layer repulsion and gravity. In the polar rotational algorithm, $T_{det}$ is the deterministic torque, which may include contributions from a mismatch in electrostatic forces near the Janus boundary and gravitational force due to a mismatch in the density of the cap and particle. $G(\Delta t)$, which has $\langle G^2 \rangle = 2D_{r,\theta}\Delta t$ variance, is Gaussian random rotation due to the stochastic torque. Electrostatic torque results from the asymmetry in electrostatic force arising when the Janus boundary separating the two hemispheres of different properties rotates towards the wall ($\theta = 90^0$). The gravitational torque depends on the polar orientation of the particle and the weight of the cap, which in turn depends on the thickness, total size, and cap material. Note that height (Eq. 14) and rotation (Eq. 15) algorithms impact each other. The height of the particle will impact rotation by affecting the rotational diffusion coefficient, electrostatic and rotational stochastic torques. Also, the orientational position of the particle will impact z-axis translational by affecting electrostatic the force between the particle and the substrate.

    A MATLAB code was developed to implement the Brownian dynamics simulation for our system. Zeta potential is a key factor in calculating electrostatic double layer repulsion force between colloidal particles and surrounding media. We addressed the challenge of accounting for variations in zeta potential with an existing meshing method [29] in which the sphere was divided into small parts in both the azimuth ($\varphi$) and polar ($\theta'$) angles. Each small meshed region has its own value of zeta potential; the small curved area was projected parallel to the boundary. The electrostatic force was calculated between each small projected flat surface and the substrate. The sum of the electrostatic interactions between the small projected area and the substrate is the double layer electrostatic force between the particle and the substrate. At each time step, this force was calculated as one of the conservative forces ($F$) in Eq. (14). Finally, the time step for each simulation was set as $\Delta t = 5$ ms. For all the system conditions studied here, the stepping algorithms were run 10 times each for $4.8 \times 10^6$ time steps. The number of observations at each separation height or orientation is the average of 10 sets of simulations. In all simulations the Janus particle

cap includes a 2.5 nm titanium as a sublayer in addition to the reported gold layer thickness. Experiments regularly include a thin layer of titanium to increase adhesion of a gold coating on a polystyrene particle. A comprehensive simulation conditions and particle properties is provided in Table I.

TABLE I. Simulation and particle conditions.

| Property Name | Value |
|---|---|
| Particle diameter | 1 μm – 6 μm |
| Particle material | Polystyrene |
| **Polystyrene Density** | 1.055 gr$^1$cm$^{-3}$ |
| Gold coating thickness | 2 nm – 20 nm |
| **Gold density** | 19.32 gr$^1$cm$^{-3}$ |
| **Titanium Density** | 4.5 gr$^1$cm$^{-3}$ |
| Temperature | 298.15 K |
| Electrolyte concentration | 1 mM |
| Time step | 5 ms |
| Initial orientation position | π/2 |
| Number of time steps | 4.8×10$^6$ |
| Surface zeta value | -50 mV |
| Particle coated side zeta value | -5 mV |
| Particle un-coated side zeta value | -50 mV |
| Number of iterations for each simulated condition | 10 |

## III. RESULTS & DISCUSSION

**A.** *Influence of deterministic torque on particle rotation.* We defined a dimensionless rotation number (DRN - $\hat{\theta}$) that balances the deterministic and stochastic torques:

$$\hat{\theta} = \frac{\Delta\theta_{det}}{\Delta\theta_{sto}} \tag{16}$$

where $\Delta\theta_{det}$ and $\Delta\theta_{sto}$ are the rotational displacements of the Janus particle in the polar direction due to deterministic and stochastic torques, respectively. This dimensionless number furnished a direct quantitative measure of the relative influence of deterministic torque as compared to stochastic torque on the particle. Stochastic rotational fluctuations dominate at small values of DRN, while deterministic torque is important at large values of DRN. We used two methods to calculate DRN. Namely, we calculated the DRN from an analytical expression and from the results of the BDS. The analytical expression for calculating the deterministic rotational displacement is:

$$\Delta\theta_{det-anl}(\Delta t) = \frac{D_{r,\theta\_Bulk} <T_{det-anl}> \Delta t}{kT} \qquad (17)$$

where $<T_{det-anl}>$ is the average deterministic torque over 180 orientations. We used the variance ($\langle G^2 \rangle = 2D_{r,\theta}\Delta t$) for calculating the random torque on the rotation of the Janus particle (Eq. 15) as the stochastic contribution. Hydrodynamic hindrance was neglected for the analytical expression to isolate the impact of each torque contribution.

Figure 2(a) shows the impact of stochastic torque embodied in the rotational displacement experienced by a Janus particle as a function of diameter and with a gold cap thickness of 20 nm. Increasing the diameter of the particle induced a decrease in the stochastic torque because of the correlation between the diameter of the Janus particle and rotational diffusion coefficient (see $G(\Delta t)$ term of Eq. (15)). Deterministic torque was calculated for the same conditions. Deterministic torque had a qualitatively similar impact on rotation as stochastic torque (see Fig. 2(b)); the magnitude of rotation decreased with increasing particle diameter at fixed cap thickness. However, deterministic torque decreased more slowly with increasing diameter. The origin of this trend for deterministic torque is in the competing effects of diameter on rotational displacements. The rotational diffusion coefficient (see Eq. 17) decreased with increasing diameter, but the magnitude of deterministic torque increased with increasing diameter due to both the growth of cap weight and displacing this weight further from the center. Figure 2(c) shows the DRN, which is the quotient of values summarized in Figs. 2(a) & 2(b). The positive slope of DRN as a function of diameter indicates that deterministic torque had an increased influence as the diameter of the particle increased.

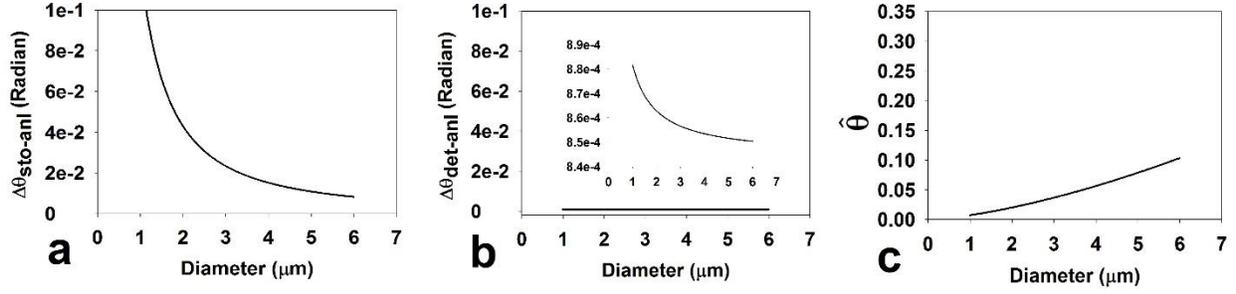

FIG. 2. Analytical calculations for the impact of (a) stochastic torque, (b) deterministic torque, and (c) the balance of these two torques via the dimensionless rotation number (DRN) on a Janus particle with a cap of 2.5 nm titanium and 20 nm gold. The inset in (b) has an amplified y-axis. These data show that although both stochastic and deterministic torque decreased with increasing diameter, deterministic torque decreased more slowly as a function of diameter. Deterministic torque, which induces rotational quenching of the Janus particle, became more important as particle size increased.

In addition to the analytical expressions summarized above, BDS was used to determine the impact of deterministic and stochastic torque on the rotation of a Janus particle. We calculated the mean polar orientation change from 10 runs of 4.8 million time steps. Figure 3(a) shows the DRN at various diameters and fixed cap thickness. In general, increasing the diameter caused an increase in the DRN, again indicating an increased importance of deterministic torque as compared to stochastic torque. These results are qualitatively similar to those obtained from the analytical solution (Fig. 2(c)). However, there are some quantitative differences between the two sets of data. First, there is a local minimum at a diameter of approximately 4 μm in results obtained from BDS as a consequence of the competing effects of rotational diffusion and cap weight, along with the inclusion of hydrodynamic hindrance in this numerical solution. Second, the magnitudes are different due to role of the random numbers in calculating BDS. The stochastic part of the DRN was calculated with BDS by taking the mean of the absolute value of orientation displacements. This value was smaller than the analytical value for stochastic fluctuations calculated from eq. 15. Given the stochastic part of the DRN is in the denominator, the resulting quotient of orientation displacements from the deterministic and stochastic parts was systemically larger for the BDS summarized in Figs. 3. Figure 3(b) shows that increasing coating thickness caused an increase in DRN. Growth of deterministic torque as a function of coating thickness is the central reason for amplification of DRN. Deterministic torque increases, while stochastic torque remains constant, as a function of coating thickness. For thicker gold coatings, Figure 3(b) predicts the dynamics of the Janus particle to be dominated by the presence of the cap.

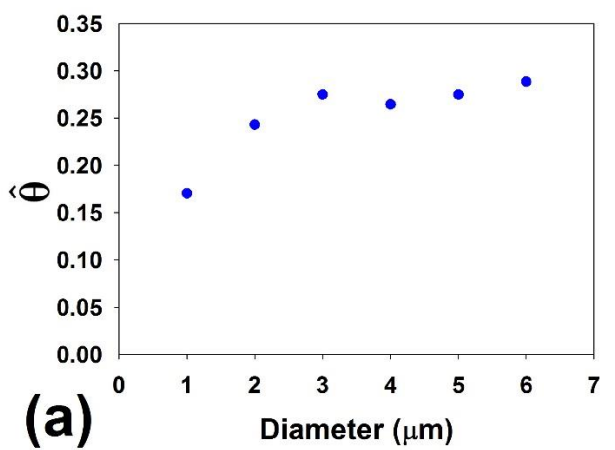

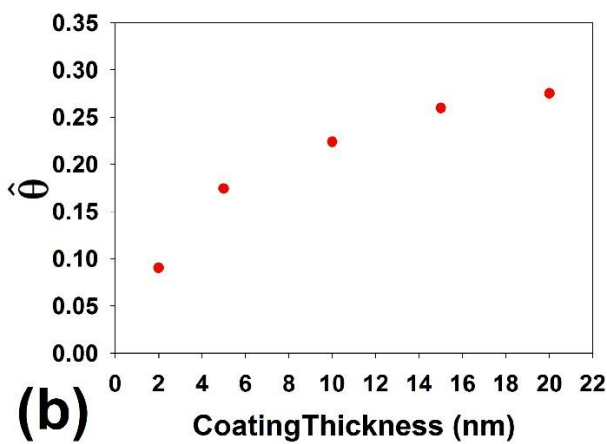

FIG. 3. (a) Effect of particle size on the DRN of Janus particles calculated with Brownian dynamics simulations (BDS) at cap of 2.5 nm titanium and 20 nm gold. (b) Effect of coating thickness on the DRN of Janus particles from BDS perspective at 3 μm particle diameter.

Results summarized in this section provide evidence that a cap will impact the dynamics of a Janus particle, with the influence of deterministic gravitational torque increasing as a function of particle diameter and cap thickness. Of note is the range over which deterministic torque will become relevant to the dynamics of a Janus particle near a boundary. For instance, Fig. 3(a) shows that increasing the particle diameter from 1 μm to 3 μm will increase DRN from 0.17 to 0.28 or from 17% to 28%. This matches the range of particle sizes often employed in experiments. Thus, when comparing measurements of Janus particle dynamics of different particle sizes, one should consider the increased importance of deterministic torque at larger particle size or thicker caps.

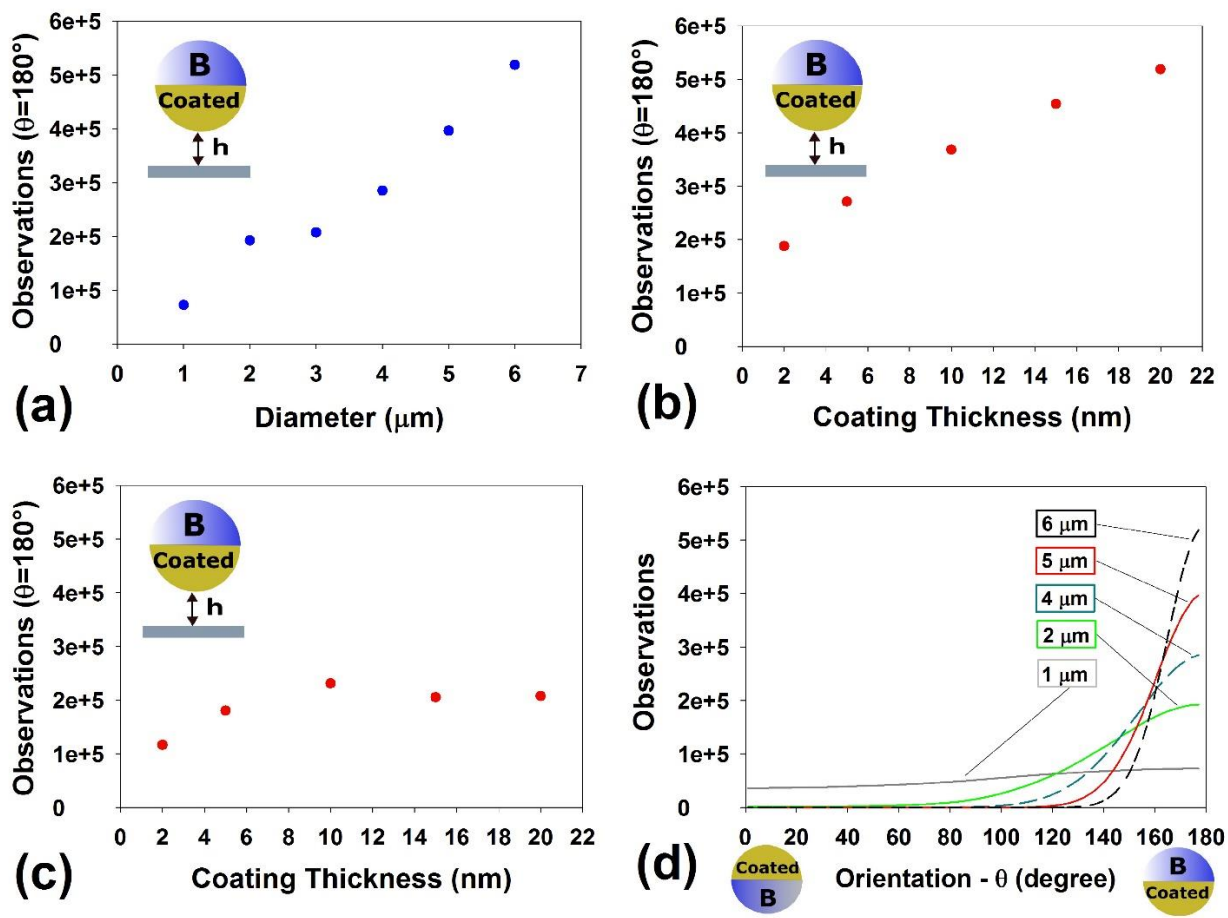

FIG. 4. (a) Impact of particle size on the number of observation for a particle with cap oriented downward (BDS - $C_{th}$ is 20 nm gold and 2.5 nm titanium). (b) Coating thickness impact on the number of observations for a particle with cap oriented downward (BDS - 6 μm). (c) Coating thickness impact on the number of observations for a particle with cap oriented downward (BDS - 3 μm). (d) Number of observations at different orientations for various particle sizes and same coating thickness (4.8 m iterations - BDS - $C_{th}$ is 20 nm gold and 2.5 nm titanium). – **B** referred to uncoated side

**B.** *Probability and potential energy landscapes for a Janus particle with cap of non-matching density.* Data summarized in the previous section suggest that deterministic torque is relevant to the dynamics of a Janus particle near a boundary. The outcome of such relevance is that the probability of finding a Janus particle at a given state will be altered by the deterministic torque. The probability density of translational and orientation states was obtained to assemble a histogram and subsequently calculate the effective interaction landscape experienced by the Janus particle under various conditions. Figures 4(a) - 4(c) summarizes data showing the impact of various parameters on the polar orientation of a Janus particle. The number of observations reported in these figures is that of a cap-down orientation, which corresponds to the 'quenched' state. Stronger quenching is associated with a larger number of observations at $\theta = 180^0$. As shown in the previous section, particle diameter at a fixed cap thickness had a strong effect on rotational quenching (see Fig. 4(a)). The number of observations at $\theta = 180^0$ increased as diameter was increased. Increasing the coating thickness increased the number of observations at $\theta = 180^0$ (see Fig. 4(b)), although note that at particle diameter equal to 3 μm (see Fig. 4(c)), the slope of the number of observations at $\theta = 180^0$ changed as the thickness increased after 10 nm. Although the growth of gravitational torque is linear with cap thickness, the weight of the particle (as the cap thickness grows) also increased. Increased particle weight brought the particle closer to the boundary, thereby decreasing the rotational diffusion coefficient. Consequently, growth in the term dictating deterministic fluctuations, which contains the product of rotational diffusion coefficient and torque (see Eq. 15), grows sub-linearly with cap thickness. Further, the reduction in rotational fluctuations is stronger with particle size in the deterministic term as compared to the stochastic term because the former contains diffusion coefficient to the power 1, while the latter has fluctuations that are proportional to the power ½. Nevertheless, both figures demonstrate the critical importance of changes in cap weight to rotational quenching of a Janus particle. Larger particle diameters and thicker caps enhance the influence of deterministic torque, thereby increasing the probability of a rotationally quenched Janus particle.

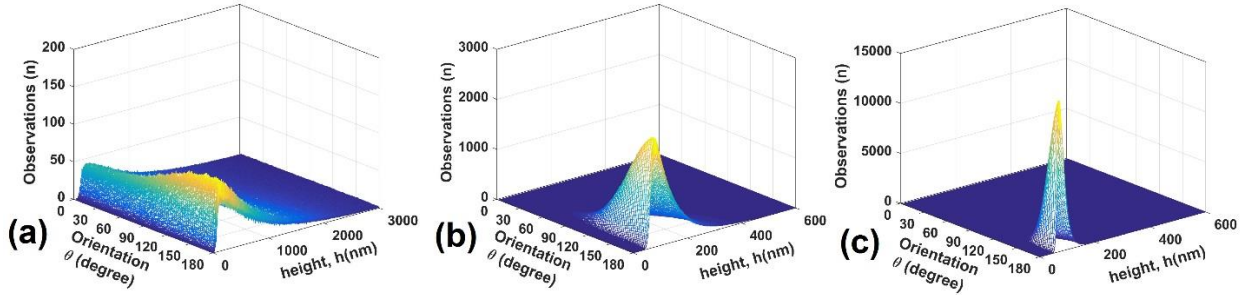

C.

D. FIG. 5. Histogram landscape for Janus particles with 2.5 nm titanium and 20 nm gold coatings at different diameters. (a) 1 μm. (b) 3 μm. (c) 6 μm. The reader should note the difference in y-axis scale.

Figure 5 shows a histogram landscape of observations in separation distance and orientation. Note the significant change in probability density for variation in particle size from 1 μm (Fig. 5(a)) to 6 μm (Fig. 5(c)) with the same coating thickness. The distribution of states spreads across all orientations for a Janus particle of 1 μm diameter and 20 nm cap thickness, meaning the Janus particle is only weakly quenched at these conditions. However, merely increasing the particle diameter to 3 μm and then 6 μm induces orientational states that are highly populated around $\theta = 180^0$. Deterministic torque became increasingly important at larger diameters due to the mismatch in the cap and particle core densities. This behavior is in contrast with the 1 μm diameter particle, which experienced random rotation such that orientational states are distributed across all available θ.

As was done previously for separation distance observations of isotropic spheres [63], observations of position and orientation were used to calculate the potential energy landscape for a Janus particle. Histogram landscapes shown in the previous section were interpreted to obtain the potential energy of interaction for a Janus particle with a cap of non-matching density. We assumed the probability of finding a Janus particle at a particular separation distance and orientation were independent and equal to the product of those individual probabilities:

$$p(h,\theta) = \underbrace{A_h e^{-\phi_h/kT}}_{p_h} \underbrace{A_\theta e^{-\phi_\theta/kT}}_{p_\theta} = A e^{-\phi_c/kT} \tag{18}$$

where $\phi_h$ is the potential energy associated with changes in separation distance, $\phi_\theta$ is the potential energy associated with changes in orientation, $\phi_c$ is the total colloidal potential energy ($\phi_c = \phi_h + \phi_\theta$), and A is a normalization constant chosen such that the cumulative probability summed over all states equals 1. Equation 18 can be rearranged and the normalization constant eliminated by subtracting the potential energy of the most probable state $\phi_c(h_m, \theta_m)$, where h$_m$ and θ$_m$ are the most probable separation distance *and* orientation corresponding to a maximum in the probability density landscape. Thus, the potential energy landscape was calculated by:

$$\frac{\phi_c(h,\theta) - \phi_c(h_m,\theta_m)}{kT} = \ln \frac{n(h_m,\theta_m)}{n(h,\theta)} \tag{19}$$

where n(h$_m$, θ$_m$) is the maximum number of particle observations among all heights and orientations, ϕ$_c$(h$_m$, θ$_m$) is the potential energy at 'most probable' position and orientation, n(h, $\theta$) and ϕ$_c$(h, $\theta$) are the number of particle observation and potential energy respectively at some specific height and polar orientation.

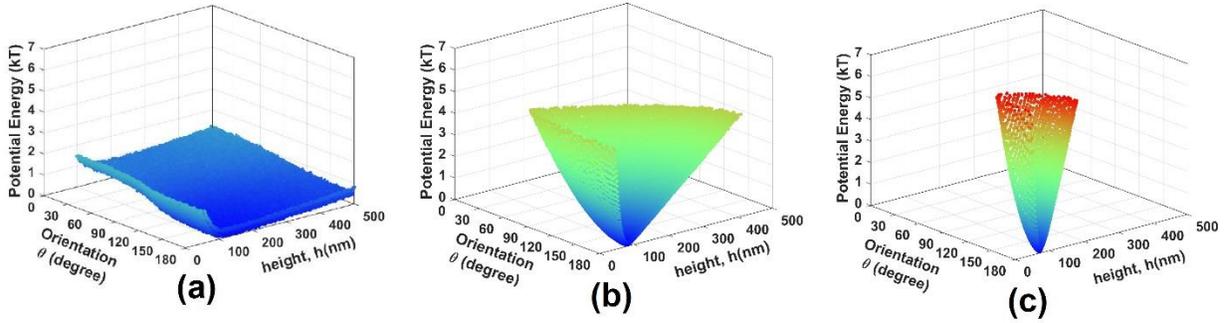

FIG. 6. (a) Potential energy landscape of 1 μm particle with 2.5 nm titanium and 20 nm gold cap. (b) Potential energy landscape of 3 μm particle with 2.5 nm titanium and 20 nm gold cap. (c) Potential energy landscape of 6 μm particle with 2.5 nm titanium and 20 nm gold cap.

Figure 6 shows the potential energy landscape for a Janus sphere of varying diameter and 20 nm gold cap. In Figs. 6(b) & 6(c), there is a minimum PE configuration belonging to the location of maximum observations among all orientations, corresponding to the coated side facing the wall. Comparing these three profiles, from a particle with diameter 1 μm to one with diameter 6 μm, illustrates the impact of a cap on conservative interactions. The potential energy landscape for a particle of 1 μm diameter nearly matches that of a Janus particle with density matching cap. Larger particle diameters with correspondingly larger gravitational torques have landscapes with a clear minimum. A colloidal particle will typically only sample states of a few kT. The projected two-dimensional view of the energy landscapes then shows how the presence of a cap will restrict the sampled states of a Janus particle. Increasing particle diameter at constant cap thickness severely limited the available orientations of the particle for a fixed energy state (see Figure 7). Although a 1 μm particle with 20 nm thick cap will sample most orientations, a 6 μm particle with the same cap thickness will sample only a fraction of possible orientations. Variations of potential energy landscapes for Janus particles of different particle size will help to explain observations of stable and unstable positions for a coated particle.

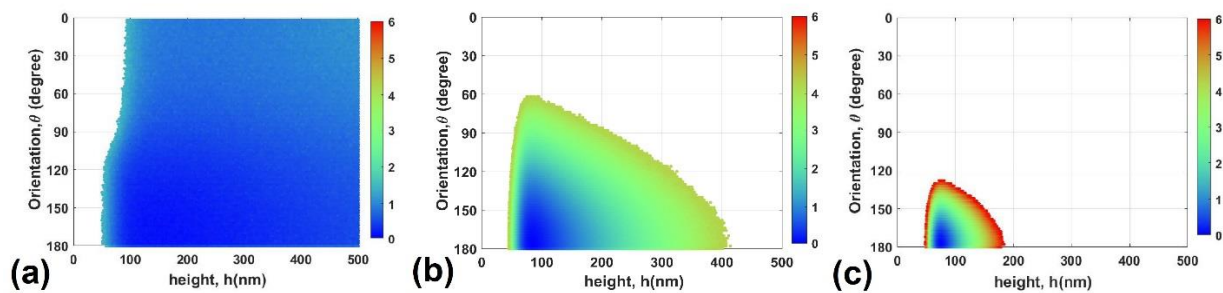

FIG. 7. (a) 2D cut away panel of BDS Potential Energy landscape of 1 μm PS with 2.5 nm titanium and 20 nm gold cap. (b) 2D cut away panel of BDS Potential Energy landscape of 3 μm PS with 2.5 nm titanium and 20 nm gold cap. (c) 2D cut away panel of BDS Potential Energy landscape of 2.5 nm titanium and 6 μm PS with 20 nm gold cap.

## IV. CONCLUSION

Brownian dynamics simulations were used to predict the rotational and translational displacements of a Janus particle with cap of non-matching density. These simulated data provided evidence that gold caps of thickness 5 nm – 20 nm on particles of diameter 1 μm to 6 μm may strongly influence the rotational dynamics of the particle. Cap-down or 'quenched' orientations arise when the balance of deterministic and stochastic torque is dominated by the former. Deterministic torque arising from the weight of the cap depended on particle size and cap thickness. Our parametric variation found that at experimentally relevant particle sizes (> 1 μm) or cap thicknesses (> 5 nm), the particle was strongly quenched such that most observations of orientation were in the cap-down state. Further, histogram landscapes were inverted to calculate the potential energy landscape for Janus particles. The energy landscapes showed that Janus particles of typical size and coating thickness will sample only a limited number of orientation states. These simulations and associated analysis revealed the importance of considering the cap weight of a Janus particle, especially when designing new materials and developing new applications that rely on particle dynamics or transport. Further, Janus particles have also been suggested as probes of local rheology and mechanics of a material. The phenomena described herein should be taken into account when utilizing Janus particles in this manner.

## V. ACKNOWLEDGEMENTS

This work was supported by the Cleveland State University Office of Research Startup Fund and the National Science Foundation CAREER Award, NSF no. 1752051. This work was supported in part by an allocation of computing time from the Ohio Supercomputer Center. SR acknowledges the support from the office of the Vice President for Research and the Provost's office at the University of Oklahoma.

**APPENDIX A: Center of Mass Calculation.**

The geometry required for the center of mass in the torque calculation was that of a semi-hollow hemisphere (Fig. 8). The axisymmetric nature of the relevant geometry allowed for the center of mass to be obtained at some position along the x-axis. Further, only an elemental cut of the cap was required to account for the unmatched torque of the Janus particle when rotated away from θ = 0° or θ = 180°. Finally, the cap thickness was small enough as compared to particle radius that it was neglected in center of mass calculations. The center of mass was calculated by:

$$CM = \frac{\int x \, dm}{M} = \frac{\int x \, dA}{\int dA} \quad (A1)$$

Where $dm$ is a differential mass element, $M$ is total mass, $dA$ is a differential surface area element, and $A$ is total surface area. By replacing $x$ and $dA$ values with the expressions found in Fig. 8, $CM$ was calculated:

$$\frac{\int_0^\alpha a\cos\psi \sin\varphi\, a\, d\psi\, a\cos\psi\, d\varphi}{\int_0^\alpha a\, d\psi\, a\cos\psi\, d\varphi} \qquad (A2).$$

The result following integration is the center of mass:

$$CM = \frac{2a}{\pi\times\sin(\alpha)} \times \left(\frac{\alpha}{2} + \frac{\sin(2\alpha)}{4}\right) \qquad (A3)$$

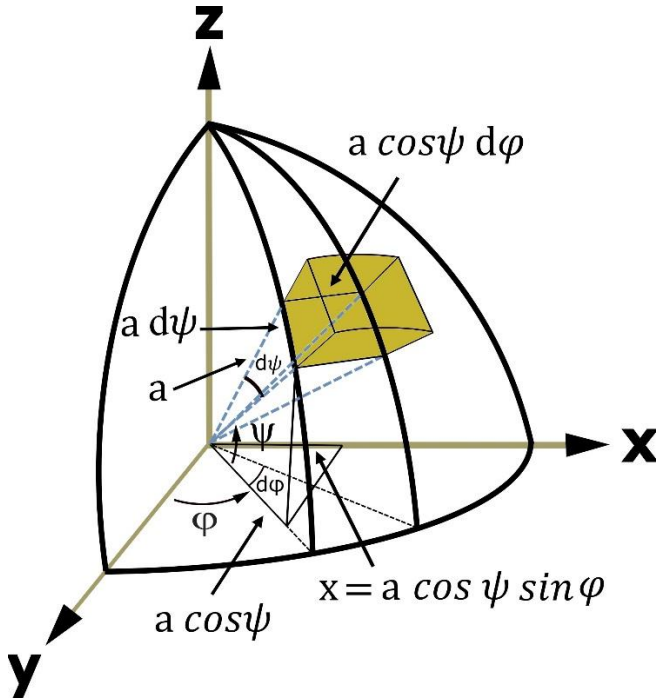

FIG. 8. A schematic of a semi-hollow hemisphere.